\documentstyle[sprocl,psfig]{article}
\begin{document}
\title{Numerical Investigation of Singularities }

\author{Beverly K. Berger}

\address{Physics Department, Oakland University,
Rochester, MI 48309 USA}

\maketitle \abstracts{Numerical exploration of the properties of singularities
could, in principle, yield detailed understanding of their nature in physically
realistic cases. Examples of numerical investigations into the formation of
naked
singularities, critical behavior in collapse, passage through the Cauchy
horizon, chaos of the Mixmaster singularity, and singularities in spatially
inhomogeneous cosmololgies are discussed.}

\section{Introduction}
The singularity theorems \cite{wald84,hawking67,he73,hp70} state that
Einstein's equations will not evolve
regular initial data arbitrarily far into the future or the past. An
obstruction
such as infinite curvature or the termination of geodesics will always arise to
stop the evolution somewhere. The simplest, physically relevant solutions
representing for example a homogeneous, isotropic universe
(Friedmann-Robertson-Walker (FRW))
or a spherically symmetric black hole (Schwarz\-schild) contain space-like
infinite curvature singularities.  Although, in principle, the presence of a
singularity
could lead to unpredictable measurements for a physically realistic
observer, this does not happen for these two
solutions. The surface of last scattering of the cosmic microwave background in
the cosmological case and the event horizon in the black hole (BH) case
effectively
hide the singularity from present day, external observers. The extent to which
this
``hidden'' singularity is generic and the types of singularities that appear in
generic
spacetimes remain major open questions in general relativity. The questions
arise quickly since other exact solutions
to Einstein's equations have singularities which are quite different
from those described above. For example,
the charged BH (Reissner-Nordstrom solution) has a time-like singularity. It
also
contains a Cauchy horizon (CH) marking the boundary of predictability of
space-like
initial data. A test observer can pass through the CH to another region of the
extended spacetime. More general cosmologies can exhibit singularity behavior
different from that in FRW. The Big Bang in FRW is classified as an
asymptotically velocity term dominated (AVTD) singularity \cite{els,IM} since
any
spatial curvature term in the Hamiltonian constraint
becomes negligible compared to the square of the expansion rate as the
singularity is approached. However, some anisotropic, homogeneous models
exhibit Mixmaster dynamics (MD) \cite{bkl71,misner69} and are not AVTD---the
influence of the spatial scalar curvature can never be neglected.

Once the simplest, exactly solvable models are left behind, understanding of
the
singularity becomes more difficult. Other chapters
\cite{moncriefgr14,rendallgr14} describe recent analytic progress. However,
such methods yield either detailed knowledge of unrealistic, simplified
(usually by
symmetries) spacetimes or powerful, general results that do not contain
details. To
overcome these limitations, one might consider numerical methods to evolve
realistic spacetimes to the point where the properties of the singularity may
be
identified. Of course, most of the effort in numerical relativity applied to BH
collisions has addressed the avoidance of singularities.\cite{finngr14} One
wishes to
keep the computational grid in the observable region outside the horizon. Much
less computational effort has focused on the nature of the singularity itself.
Numerical calculations, even more than analytic ones, require finite values for
all
quantities. Ideally then, one must describe the singularity by the asymptotic
non-singular
approach to it. A numerical method which can follow the evolution into
this asymptotic regime will then yield information about the singularity. Since
the
numerical study must begin with a particular set of initial data, the results
can
never have the force of mathematical proof. One may hope, however, that such
studies will provide an understanding of the ``phenomenology'' of singularities
that will eventually guide and motivate rigorous results.

In the following, we shall consider examples of numerical study of
singularities
both for asymptotically flat (AF) spacetimes and for cosmological models. These
examples have been chosen to illustrate primarily numerical studies whose focus
is the nature of the singularity itself. In the AF context, we shall consider
two
questions.
The first is whether or not naked singularities exist for realistic matter
sources. One approach is to explore
highly non-spherical collapse looking for spindle or pancake singularities. If
the
formation of an event horizon requires a limit on the aspect ratio of the
matter,\cite{thorne74} such configurations may yield a naked singularity.
Another
approach is to probe the limits between initial configurations which lead to
black
holes and those which yield no singularity at all (i.e.~flat spacetime
plus radiation) to explore the singularity as the BH mass goes to zero. This
quest
led naturally to the discovery of critical
behavior in the collapse of a scalar field.\cite{choptuik93} The other
question which is now beginning to yield to
numerical attack involves the stability of the Cauchy horizon in charged
or rotating black holes where it has been conjectured \cite{wald84,ch82} that a
real
observer, as opposed to a test mass, cannot pass through the CH since realistic
perturbed spacetimes will convert the CH to a true singularity.
In cosmology, we shall consider both the behavior of the Mixmaster model
and the issue of whether or not its properties are applicable to generic
cosmological singularities.  Although numerical evolution of the Mixmaster
equations
has a long history, recent developments are motivated by inconsistencies
between
the known sensitivity to initial conditions and standard measures of the chaos
usually associated with such
behavior.\cite{moser73,rugh90,berger94,francisco88,burd90,hobill91,pullin91}
Belinskii, Khalatnikov, and Lifshitz (BKL) long ago claimed \cite{bkl71} that
it is
possible to formulate the generic cosmological solution to Einstein's equations
near the singularity as a Mixmaster universe at every point. While others have
questioned the the validity of this claim,\cite{bt79} there is as yet no
evidence
either way for such behavior in spatially inhomogeneous cosmologies. We shall
discuss a numerical program to address this issue.

\section{Singularities in AF spacetimes}
\subsection{Naked singularities and the hoop conjecture}
The strong cosmic censorship conjecture \cite{penrose69} requires a
singularity formed from regular, asymptotically flat initial data to be
hidden from an external observer by an event horizon. Counter examples
have been known for a long time but tend to be dismissed
as unrealistic in some way. The goal of a numerical approach is to try to
search for naked singularities arising from physically reasonable initial
conditions. A possible regime for such systems is motivated by Thorne's
``hoop conjecture'' \cite{thorne74} that collapse will yield a black hole
only if a mass $M$ is compressed to a region with circumference $C \le
4 \pi M$ in all directions. (Note that one must take care to
define $C$ and $M$ especially if the initial data are not at least axially
symmetric.)
If the hoop conjecture is true, naked singularities may form
if collapse can yield $C \ge 4\pi M$ in some direction.
The existence of a naked singularity is inferred from the absence of an
apparent horizon (AH) which can be identified locally by
following null geodesics. Although a definitive identification of a naked
singularity
requires the event horizon (EH) to be proven to be absent, to identify an EH
requires knowledge of the entire spacetime. Methods to find an EH in a
numerically
determined spacetime have only recently become available and have not been
applied
to this issue.\cite{seidel95}
To attempt to produce naked singularities, Shapiro and Teukolsky (ST)
\cite{shapiro91} considered collapse of prolate spheroids of collisionless
gas. (Nakamura and Sa\-to~\cite{nakamura82} had previously
studied the collapse of
non-rotating deformed stars with an initial large reduction of internal
energy and apparently found spindle or pancake singularities
in extreme cases.)
ST solved the general relativistic Vlasov equation for
the particles along with Einstein's equations for the gravitational field.
Null geodesics were followed to identify an AH if present.
The curvature invariant $I =R_{\mu \nu \rho \sigma}R^{\mu \nu \rho \sigma}$
was also computed. They found that an AH (and presumably a BH) formed if
$C \le 4\pi M < 1$ everywhere but no AH (and presumably a naked singularity)
in the opposite case. In the latter case, the evolution (not surprisingly)
could not proceed past the moment of formation of the singularity.
In a subsequent study, ST \cite{shapiro92} also showed that a
small amount of rotation (counter rotating particles with no
net angular momentum) does not prevent the formation of a naked
spindle singularity. However, Wald and Iyer
\cite{wald91} have shown
that the Schwarzschild solution has a time slicing whose evolution approaches
arbitrarily close to the singularity with no AH in any slice (but,of course,
with an EH in the spacetime). This may mean that there is a
chance that the increasing prolateness found by ST in effect changes the
slicing to
one with no apparent horizon just at the point required by the hoop conjecture.
While, on the face of it, this seems unlikely, Tod gives an example where
the AH does not form on a chosen constant time slice---but rather different
portions form at different times. He argues that a numerical simulation might
be
forced by the singularity to end before the formation of the AH is complete.
Such an AH would not be found by the simulations.\cite{tod92} In response,
Shapiro and Teukolsky considered equilibrium sequences of prolate
relativistic star clusters.\cite{shapiro93} The idea is to counter the
possibility that an EH might
form after the simulation must stop. If an equilibrium configuration is
non-singular, it cannot contain an EH since singularity theorems say that
an EH implies a singularity. However, a sequence of non-singular equilibria
with rising $I$ ever closer to the spindle singularity would lend support
to the existence of a naked spindle singularity since one can approach
the singular state without formation of an EH. They constructed this
sequence and found that the singular end points were very similar to
their dynamical spindle singularity.

Motivated by ST's results,\cite{shapiro91} Echeverria \cite{echeverria93}
numerically
studied the properties of the naked singularity that is known to form in the
collapse of an infinite, cylindrical dust shell.\cite{thorne74} While
the asymptotic state can be found analytically, the approach to it must
be followed numerically. The analytic asymptotic solution can be matched
to the numerical one (which cannot be followed all the way to the collapse)
to show that the singularity is strong (an observer experiences infinite
stretching parallel to the symmetry axis and squeezing perpendicular to
the symmetry axis). A burst of gravitational radiation
emitted just prior to the formation of the singularity
can be studied.

A very recent numerical study of the hoop conjecture was made by Chiba
et al.\cite{chiba94} Rather than a dynamical collapse model, they searched
for AH's in analytic initial data for discs, annuli, and rings. Previous
studies of this type were done by Nakamura et al \cite{nakamura88} with oblate
and prolate spheroids and by Wojtkiewicz \cite{wojtkewicz90} with
axisymmetric singular lines and rings.
The summary of their results is that an AH forms if $C \le 4\pi M \le
1.26$. (Analytic results due to Barrab\`{e}s et al
\cite{barrabes91,barrabes92} and Tod \cite{tod92} give similar quantitative
results with different initial data classes and (possibly) definition
of $C$.) The results of all these searches for naked singularities are
controversial
but could be resolved if the presence or absence of the EH could be determined.

\subsection{Critical behavior in collapse}
We now consider an effect that was originally found by Choptuik
\cite{choptuik93} in a numerical study of the collapse of a spherically
symmetric massless scalar field.
We note that this is the first completely new phenomenon in general relativity
to be discovered by numerical simulation. In collapse of a scalar field,
essentially two things can happen: either
a BH forms or the scalar waves pass through each other and disperse.
Choptuik discovered that for any 1-parameter set of initial data labeled by
$p$, there is a critical value $p^{\ast}$ such that $p > p^{\ast}$ yields
a BH. He found
\begin{equation}
\label{chopscale}
M_{BH} \approx C_F \left( p - p^{\ast} \right)^{\gamma}
\end{equation}
where $M_{BH}$ is the mass of the eventual BH. The constant $C_F$ depends
on the parameter of the initial data that is selected but $\gamma \approx .37$
is the
same
for all choices. Furthermore, in terms of logarithmic variables
$\rho = \ln r + \kappa$, $\tau = \ln (T_0^{\ast} - T_0) + \kappa$
($T_0$ is the proper time of an observer at $r = 0$ with $T_0^{\ast}$
the finite proper time at which the critical evolution concludes), the
waveform $X$ repeats (echoes)  at intervals of $\Delta$ in $\tau$ if
$\rho$ is
rescaled to $\rho - \Delta$, i.e. $X(\rho - \Delta, \tau - \Delta)
\approx X(\rho, \tau)$. The scaling behavior (\ref{chopscale}) demonstrates
that the minimum BH mass (for
bosons) is zero. The critical solution itself is a
counter-example to cosmic censorship (since the
formation of the zero mass BH causes high curvature regions
become visible at $r = \infty$).  (See, e.g., the discussion in Hirschmann
and Eardley.\cite{hirschmann95a})

Soon after this discovery, scaling and critical phenomena were found in
a variety of contexts. Abrahams and Evans \cite{abrahams93} discovered the
same phenomenon in axisymmetric gravitational wave collapse with a different
value of $\Delta$ and, to within numerical error, the same value of $\gamma$.
(Note that the rescaling of $r$ with $e^{\Delta} \approx 30$ required
Choptuik to use adaptive mesh refinement (AMR) to distinguish subsequent
echoes. Abrahams and
Evans' smaller $\Delta$ ($e^{\Delta} \approx 1.8$) allowed them to see
echoing with their $2+1$ code without AMR.) Recently,
Gar\-fin\-kle~\cite{garfinkle95} has confirmed Choptuik's results with a
completely
different algorithm that does not require AMR. His use of Goldwirth and
Piran's \cite{goldwirth87} method of simulating Christodoulou's
\cite{christodoulou87} formulation of the spherically symmetric scalar field in
null
coordinates allowed the grid
to be automatically rescaled by choosing the edge of the grid to be the
null ray that just hits the central observer at the end of the critical
evolution. (Missing points of null rays that cross the central observer's
world line are replaced by interpolation between those that remain.)
Hamad\'{e} and Stewart \cite{hamade95} have also repeated Choptuik's
calculation using null coordinates and AMR. They are able to achieve greater
accuracy and find $\gamma = .374$.

Evans and Coleman \cite{evans94} realized that self-similar rather than
self-periodic collapse might be more tractable both numerically (since
ODE's rather than PDE's are involved) and analytically. They discovered
that a collapsing radiation fluid had that desirable property. (Note that
self-similarity (homothetic motion) is incompatible with AF. However, most
of the action occurs in the center so that a match of the self-similar
inner region to an outer AF one should always be possible.) In a series of
papers, Hirschmann and Eard\-ley~\cite{hirschmann95a,hirschmann95b} developed
a (numerical) self-similar solution to the spherically symmetric {\it complex}
scalar field equations. These are ODE's with too many boundary conditions
causing a solution to exist only for certain fixed values of $\Delta$.
Numerical
solution of this eigenvalue problem allows very accurate determination of
$\Delta$.
The self-similarity also allows accurate
calculation of $\gamma$ as follows:
The critical $p = p^{\ast}$ solution is unstable to a small change in $p$. At
any
time $t$ (where $t < 0$ is
increasing toward zero),
the amplitude $a$ of the perturbation exhibits power law growth:
\begin{equation}
\label{eardley1}
a \propto (p - p^{\ast}) (-t)^{- \kappa}
\end{equation}
where $\kappa > 0$. At any fixed $t$, larger $a$ implies larger $M_{BH}$.
Equivalently, any fixed amplitude $a = \delta$ will be reached faster for
larger eventual $M_{BH}$. Scaling arguments give the dependence of
$M_{BH}$ on the time at which any fixed amplitude is reached:
\begin{equation}
\label{eardley2}
M_{BH} \propto (-t_1)
\end{equation}
where
\begin{equation}
\label{eardley3}
(p - p^{\ast}) (-t_1)^{\kappa} \propto \delta .
\end{equation}
Thus
\begin{equation}
\label{eardley4}
M_{BH} \propto (p - p^{\ast})^{1/\kappa} .
\end{equation}
Therefore, one need only identify the growth rate of the unstable mode to
obtain an accurate value of $\gamma = 1 / \kappa$. It is not necessary to
undertake the entire dynamical evolution or probe the space of initial
data. Hirschmann and Eardley obtain $\gamma = 0.387106$ for the complex
scalar field solution
while Koiki et al~\cite{koiki95} obtain $\gamma = 0.35580192$ for the
Evans-Coleman solution. Although the similarities among the critical
exponents $\gamma$ in the collapse computations suggested a universal value,
Maison \cite{maison95} used these same scaling-perturbation methods to show
that
$\gamma$ depends on the equation of state $p = k \rho$ of the fluid in the
Evans-Coleman solution. Very recently, Gundlach \cite{gundlach95} used a
similar approach
to locate Choptuik's critical solution accurately.
This is much harder due to its discrete self-similarity. He reformulates
the model as nonlinear hyperbolic
boundary value problem with eigenvalue $\Delta$ and finds $\Delta
= 3.4439$.  As with the self-similar solutions described above, the critical
solution is found directly without the need to perform a dynamical evolution
or explore the space of initial data.
Perturbation of this solution could yield an accurate value of $\gamma$.

\subsection{Nature of the singularity in charged or rotating black holes}
Unlike the simple singularity structure of the Schwarzschild solution, where
the
event horizon
encloses a spacelike singularity at $r=0$, charged and/or rotating BH's
have a much richer singularity structure. The extended spacetimes have an inner
Cauchy horizon (CH) which is the boundary of predictability. To the future
of the CH lies a timelike (ring) singularity.\cite{wald84}
Poisson and Israel \cite{poisson89,poisson90} began an analytic study of
the effect of perturbations on the CH. Their goal was to check conjectures
that the blue-shifted
infalling radiation during collapse would convert the CH into a true
singularity and thus prevent an observer's passage into the rest of the
extended regions. By including both ingoing and back-scattered
outgoing radiation, they
find for the Reissner-Nordstrom (RN) solution that the mass function
(qualitatively
$R_{\alpha \beta \gamma \delta} \propto M / r^3$) diverges at the CH
(mass inflation). However, Ori showed both for RN and Kerr
\cite{ori91,ori92} that the metric perturbations are finite (even though
$R_{\mu \nu \rho \sigma}R^{\mu \nu \rho \sigma}$ diverges) so that an
observer would not be destroyed by tidal forces
(the tidal distortion would be finite)
and could survive passage through the CH. A numerical solution of the
Einstein-Maxwell-scalar field equations could test these perturbative
results.

Gnedin and Gnedin \cite{gnedin93} have numerically evolved the
spherically symmetric Einstein-Maxwell with massless scalar field equations
in a $2+2$ formulation. The initial conditions place a scalar field on part of
the
RN event horizon (with zero field on the rest). An asymptotically null or
spacelike singularity whose shape depends on the strength of the initial
perturbation replaces the CH. For a sufficiently strong perturbation, the
singularity is Schwarzschild-like. Although they claim to have found that
the CH evolved to become a spacelike singularity, the diagrams in
their paper show at least part of the final singularity to be null or
asymptotically
null in most cases.

Brady and Smith \cite{brady95} used the Goldwirth-Piran formulation
\cite{goldwirth87} to study the same problem. They assume the spacetime
is RN for $v < v_0$. They follow the evolution of the CH
into a null singularity, demonstrate mass inflation, and support (with
observed exponential decay of the metric component $g$) the validity of
previous analytic results \cite{poisson89,poisson90,ori91,ori92} including the
``weak'' nature of the singularity that forms. They find that the
observer hits the null CH singularity before falling into the curvature
singularity at $r = 0$. Whether or not these results are in conflict with
Gnedin and Gnedin \cite{gnedin93} is unclear.\cite{bonanno}

We finally mention H\"{u}bner's \cite{huebner94} numerical scheme to evolve
on a conformal compactified grid using Friedrich's formalism.\cite{Fried} He
considers the spherically symmetric
scalar field model in a $2+2$ formulation. So far, this code has been used
to locate singularities and to identify Choptuik's scaling.\cite{choptuik93}

\section{Singularities in cosmological models}
\subsection{Mixmaster dynamics}

Belinskii, Khalatnikov, and Lifshitz \cite{bkl71} (BKL) described the
singularity approach of
vacuum Bianchi IX cosmologies as an infinite sequence of
Kasner epochs whose indices change when the scalar curvature
terms in Einstein's equations become important. They were able to describe the
dynamics approximately by a map evolving a discrete set of parameters from one
Kasner epoch to the next.\cite{bkl71,cb83} For example, the Kasner indices for
the power law dependence of the anisotropic scale factors can be parametrized
by a
single variable $u \ge 1$. BKL determined that
\begin{equation}
\label{umap}
u_{n+1}=\left\{ \matrix{u_n-1 \quad \quad,\quad 2\le u_n\hfill\cr
  (u_n-1)^{-1}\quad,\quad1\le u_n\le 2\hfill\cr} \right. \quad .
\end{equation}
The subtraction in the denominator
for $1 \le u_n \le 2$ yields the sensitivity to initial conditions
associated with Mixmaster dynamics (MD).
Misner \cite{misner69} described the same
behavior in terms of the model's volume and anisotropic
shears. A multiple of the scalar curvature
acts as an outward moving potential in
the anisotropy plane.  Kasner epochs become straight line
trajectories moving out a potential corner while bouncing from one side to the
other.
A change of corner ends a BKL era when $u \to (u-1)^{-1}$.
Numerical evolution of Einstein's equations was used to
explore the accuracy of the BKL map as a descriptor of the dynamics as well as
the
implications of the map.\cite{moser73,rugh90,berger94}

Later, the BKL sensitivity to initial conditions was
discussed in the language of chaos.\cite{barrow82,khalatnikov85}
However, the chaotic nature of Mixmaster dynamics was questioned when
numerical evolution of the
Mixmaster equations yielded zero Lyapunov exponents
(LE's).\cite{francisco88,burd90,hobill91} (The LE
measures the divergence of initially nearby trajectories. Only an exponential
divergence, characteristic of a chaotic system, will yield positive exponent.)
Other numerical studies yielded positive LE.\cite{pullin91}
This issue was resolved when the LE was shown numerically and analytically
to depend on the choice of time variable.\cite{berger91,ferraz91}
Although MD itself is well-understood, its characterization as chaotic
or not is still controversial.\cite{hobillbook} Recently, Le Blanc et al
\cite{LKW} have shown (analytically and numerically)
that MD can arise in Bianchi VI$_0$
models with magnetic fields. In essence, the magnetic field provides the wall
needed to close the potential in a way that yields the BKL map for $u$.
\cite{BKB95}

There are some recent numerical studies of Mixmaster dynamics in other
theories of gravity. For example, Carretero-Gonzalez et al
\cite{carretero94} find
evidence of chaotic behavior in Bianchi IX-Brans-Dicke solutions while
Cotsakis et al \cite{cotsakis93} have shown that
Bianchi IX models in 4th order gravity theories
have stable non-chaotic solutions.

\subsection{Inhomogeneous Cosmologies}
In the remainder of this chapter, I shall discuss results of a joint project
with V.
Moncrief, D. Garfinkle, and B. Grubi\u{s}i\'{c} to explore the nature of the
generic cosmological singularity.\cite{bkbvm,bggm95}
BKL have conjectured that one should expect such a
singularity to be locally of the Mixmaster type.\cite{bkl71} The main
difficulty with the acceptance of this conjecture has been the controversy over
whether the required time slicing can be constructed globally.\cite{bt79}
Montani,\cite{montani} Belinskii,{\cite{belinskii} and
Kochnev and Kir\-il\-lov~\cite{KK} have pointed out that if the BKL conjecture
is
correct, the spatial structure of the singularity could become extremely
complicated as bounces occur at different locations at different times.  The
simplest
cosmological models which might have local MD are vacuum universes on $T^3
\times R$ with a $U(1)$ symmetry.\cite{Moncrief86} The non-commuting
Killing vectors of local MD can be constructed since only one Killing vector is
already present. The two commuting Killing vectors of the even simpler plane
symmetric
Gowdy cosmologies \cite{gowdy,bkb74} preclude their use to test the conjecture.
However, these models are interesting in their own right since they have been
conjectured to possess an AVTD singularity.\cite{bm1}

We employ a symplectic partial differential equation solver,\cite{fleck,vm83}
a type of operator splitting which singles out the AVTD limit
if it is present. If the
model is, in fact, AVTD, the approximation in the numerical scheme should
become more accurate as the singularity is approached. An outline of the method
follows.

For a field $q(x,t)$ and its conjugate momentum $\pi(x,t)$ split
the Hamiltonian operator into kinetic and potential energy subhamiltonians.
Thus,
\begin{equation}
H=\int {dx\left\{ {{\textstyle{1 \over 2}}\pi ^2+V[q]}
\right\}}=H_1(\pi )+H_2(q) .
\end{equation}
If the vector $X = (\pi,q)$
defines the variables at time $t$, then the time evolution is given by
\begin{equation}
{{dX} \over {dt}}=\{H,X\}_{PB}\equiv AX
\end{equation}
where $\{ \quad \}_{PB}$ is the Poisson bracket. The usual exponentiation
yields an evolution operator
\begin{equation}
\label{evapprox}
e^{A\Delta t}=e^{A_1(\Delta t/ 2)}e^{A_2\Delta
t}e^{A_1(\Delta t/2)}+O(\Delta t^3)
\end{equation}
for $A = A_1 + A_2$ the generator of the time evolution.
Higher order accuracy may be obtained by a better
approximation to the evolution operator.\cite{suzuki}
This method is useful when exact solutions for the
subhamiltonians are known. For the given $H$, variation of $H_1$ yields the
solution
\begin{equation}
q=q_0+\pi _0\,\Delta t\quad,\quad\pi =\pi _0,
\end{equation}
while that of $H_2$ yields
\begin{equation}
q=q_0\quad,\quad\pi =\pi _0-{{\delta V}
\over {\delta q}}\Delta t
\end{equation}
where ${{\delta V} / {\delta q}}$ is the appropriate functional derivative.
Note that $H_2$ is exactly solvable for any potential $V$ no matter how
complicated although the required
differenced form of the potential gradient may be non-trivial.
One evolves from $t$  to  $t + \Delta t$ using the exact
solutions to the subhamiltonians according to the
prescription given by the approximate evolution operator (\ref{evapprox}).

The Gowdy model on $T^3 \times R$ serves as an excellent laboratory for this
method.
It is described by the metric \cite{gowdy}
\begin{eqnarray}
\label{gowdymetric}
ds^2&=&e^{{{-\lambda } \mathord{\left/ {\vphantom {{-\lambda
} 2}} \right. \kern-\nulldelimiterspace} 2}}e^{{\tau  \mathord{\left/
{\vphantom {\tau  2}} \right. \kern-\nulldelimiterspace} 2}}(-\,e^{-2\tau
}\,d\tau ^2+d\theta ^2)\nonumber \\
 &  &+e^{-\tau }\,[e^Pd\sigma ^2+2e^PQ\,d\sigma \,d\delta +(e^PQ^2+e^{-
P})\,d\delta ^2]
\end{eqnarray}
where $\lambda$, $P$, $Q$ are functions of $\theta$, $\tau$. We impose
$T^3$ spatial topology by requiring $0 \le \theta, \sigma, \delta \le 2 \pi$
and the metric functions to be periodic in $\theta$.
If we assume $P$ and $Q$ to be small, we find them to be respectively the
amplitudes of the $+$ and $\times$ polarizations of the gravitational
waves  with $\lambda$ describing the background in which they
propagate.  The time variable $\tau$ measures the area in the symmetry
plane with $\tau = \infty$ a curvature singularity. Einstein's equations split
into two groups. The first is nonlinearly coupled wave equations for $P$
and $Q$ (where $,_a = \partial / {\partial a}$):
\begin{eqnarray}
\label{gowdywave}
P,_{\tau \tau }-\;e^{-\kern 1pt2 \tau }P,_{\theta \theta }-e^{2P}\left(
{Q,_\tau ^2-\;e^{-\kern 1pt2\tau }Q,_\theta ^2} \right)&=&0,\\
  Q,_{\tau \tau }-\;e^{-\kern 1pt2\tau }Q,_{\theta \theta }+\,2\,\left(
{P,_\tau Q,_\tau ^{}-\;e^{-\kern 1pt2\tau }P,_\theta Q,_\theta ^{}}
\right)&=&0.
\end{eqnarray}
The second contains the Hamiltonian and $\theta$-momentum constraints
respectively
which can be expressed as first order equations for $\lambda$ in terms of
$P$ and $Q$:
\begin{equation}
\label{gowdyh0}
\lambda ,_\tau -\;[P,_\tau ^2+\;e^{-2\tau }P,_\theta ^2+\;e^{2P}(Q,_\tau
^2+\;e^{-2\tau }\,Q,_\theta ^2)]=0,
\end{equation}
\begin{equation}
\label{gowdyhq}
\lambda ,_\theta -\;2(P,_\theta P,_\tau +\;e^{2P}Q,_\theta Q,_\tau )=0.
\end{equation}
This break into dynamical and constraint equations removes two of the
most problematical areas of numerical relativity from this model.  (1) The
normally difficult initial value problem becomes trivial since $P$, $Q$
and their first time derivatives may be specified arbitrarily (as long as the
total $\theta$ momentum in the waves vanishes).  (2) The constraints,
while guaranteed to be preserved in an analytic evolution by the Bianchi
identities, are not automatically preserved in a numerical evolution with
Einstein's equations in differenced form.  However, in the Gowdy model,
the constraints are trivial since $\lambda$ may be constructed from the
numerically determined $P$ and $Q$.
For the special case of the polarized Gowdy model ($Q=0$), $P$ satisfies
a linear wave equation whose exact solution is well-known.\cite{bkb74}
For this case, it has been proven that the singularity is AVTD.\cite{IM}
This has also been conjectured to be true for generic Gowdy models.\cite{bm1}

The wave
equations (\ref{gowdywave}) can be obtained by variation of the Hamiltonian
\begin{eqnarray}
\label{gowdywaveh}
H&=&{1 \over 2}\int\limits_0^{2\pi } {d\theta
\,\left[ {\pi _P^2+\kern 1pt\,e^{-2P}\pi _Q^2} \right]}\nonumber \\
  &+&{1 \over 2}\int\limits_0^{2\pi } {d\theta \,\left[ {e^{-
2\tau }\left( {P,_\theta ^2+\;e^{2P}Q,_\theta ^2} \right)}
\right]}=H_1+H_2
\end{eqnarray}
so that implementation of the symplectic method is straightforward.
Again equations obtained from the separate variations of $H_1$ and
$H_2$ are exactly solvable.  Variation of $H_1$ yields the terms in
(\ref{gowdywave}) containing time derivatives.  These have the exact
(AVTD) solution
\begin{eqnarray}
\label{avtdeq}
P&=&-\beta \tau +\ln [\alpha \,(1+\zeta ^2e^{2\beta \tau })]\to \beta
\tau \quad {\rm as} \ \tau \to \infty ,\nonumber \\
Q&=&-\;{{\zeta \,e^{2\beta \tau }} \over {\alpha \,(1+\zeta
^2e^{2\beta \tau })}}+\xi \quad\to Q_0\ \;{\rm as} \ \tau \to \infty
\ ,\nonumber \\
\pi_P &=& {-{\beta (1-\zeta^2 e^{2 \beta \tau})} \over {(1+\zeta^2 e^{2 \beta
\tau})} } \quad \to \beta \ \; {\rm as} \ \tau \to \infty \ , \nonumber \\
\pi_Q &=& -2 \alpha \beta \zeta
\end{eqnarray}
in terms of four constants $\alpha$, $\beta$, $\zeta$, and $\xi$.
To employ the AVTD solution in the symplectic method, the values of
$P$, $Q$, $\pi_P$, and $\pi_Q$ at $\tau^j$ are used to find $\alpha$, $\beta$,
$\zeta$, and $\xi$.  These are substituted in (\ref{avtdeq}) to evolve to new
values at $\tau^{j+1}$ according to (\ref{evapprox}).
Evolution with $H_2$ is still easy because $P$ and $Q$ are constant. For
completeness, we give the (2nd order) differenced form of $H_2$ as
\begin{equation}
\label{h2diffgowdy}
H_2={{e^{-2\tau }} \over {(\Delta \theta
)^2}}\sum\limits_{i=0}^N { {\left[ {\left(
{P_i-P_{i-1}} \right)^{2}+e^{P_i+P_{i-1}}\left( {Q_i-Q_{i-
1}} \right)^2} \right]} .}
\end{equation}
The exponential prefactor $e^{-2 \tau}$
in $H_2$ makes plausible the conjectured AVTD singularity.  However, $P
\to v \tau$ (for $v > 0$) as $\tau \to \infty$, (where from (\ref{avtdeq})
$v = \beta$). If $v > 1$, the term $e^{- 2 \tau} e^{2 P} Q,_{\theta}^2$ in
(\ref{gowdywaveh}) can grow rather than decay as $\tau \to \infty$. This
has led to the conjecture that the AVTD limit requires $v < 1$ everywhere
except, perhaps, at isolated values of $\theta$.\cite{bm1}

Our numerical results \cite{bkbvm,bggm95} use the
(standard) initial data $P = 0$, $\pi_P = v_0
\cos \theta$, $Q = \cos \theta$, and $\pi_Q = 0$.  This model is actually
generic for the following reasons:  The $\cos \theta$ dependence is the
smoothest nontrivial possibility.  With $\cos n \theta$, the solution is
repeated $n$ times on the grid yielding the same result with
poorer resolution.  The amplitude of $Q$
is irrelevant since the Hamiltonian (\ref{gowdywaveh}) is invariant under
$Q \to \rho Q$, $P \to P - \ln \rho$.  This also means that any unpolarized
model is qualitatively different from a polarized ($Q = 0$) one no matter
how small $Q$ is.

The accuracy and stability of the code easily allow verification of the
conjectured AVTD behavior.\cite{bkbvm} A plot of the maximum value of
$v$ vs $\tau$ (Fig.~\ref{vmax}) shows strong support for the conjecture
that $v < 1$ in the AVTD regime.
\begin{figure}[bth]
\setlength{\unitlength}{1cm}
\makebox[11.7cm]{\psfig{file=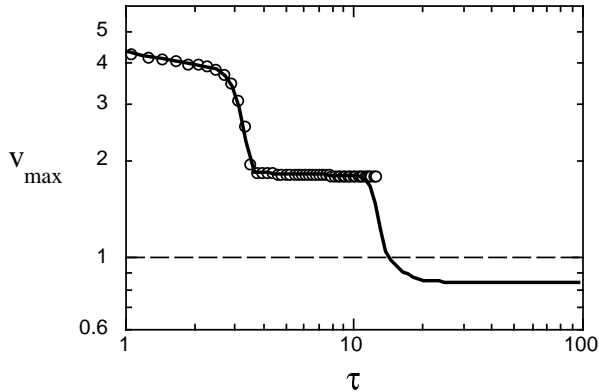,width=8cm}}
\caption[$v_{max}$ vs $\tau$]
{\protect \label{vmax}
Plot of $v_{max}$ vs $\tau$. The maximum value of $v$ is found for two
simulations with 3200 (solid line) and 20000 spatial grid points (circles)
respectively. The horizontal line indicates $v = 1$.}
\end{figure}
However, Fig.~\ref{vmax} also shows that a simulation at higher spatial
resolution begins
to diverge from one at lower resolution. Normally, failure of convergence
signals numerical problems. Here
something different is occurring. The evolution of spatial structure in $P$
depends on competition between the two nonlinear terms in the $P$ wave
equation. Approximating the wave equation by $P,_{\tau \tau} +
{\rm \ either \  of\  the\  nonlinear\  terms} = 0$ yields a first integral.
The two potentials are $V_1 = \pi_Q^2 e^{-2P}$ and $V_2 =
e^{-2 \tau} e^{2P} Q,_{\theta}^2$.  Non-generic behavior can arise at
isolated points where either $Q,_{\theta}$ or $\pi_Q$ vanishes. Say such a
point is $\theta_0$. The finer the grid, the closer will be some grid point
to $\theta_0$. Thus non-generic behavior will become more visible on a finer
grid. Detailed examination shows that the differences seen in Fig.~\ref{vmax}
are due
to the slower decay of $v$ to a value below unity at isolated grid points in
the higher resolution simulation.

Details of the high resolution simulation are shown in Fig.~\ref{gdetail}.
\begin{figure}[bth]
\setlength{\unitlength}{1cm}
\makebox[11.7cm]{\psfig{file=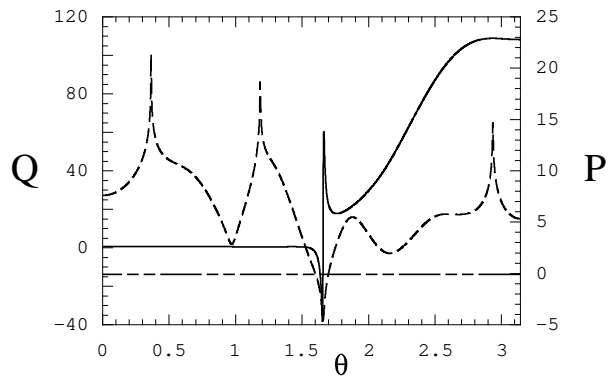,width=8cm}}
\caption
{\protect \label{gdetail}
$P$ (dashed line) and $Q$ (solid line) vs $\theta$ at $\tau = 12.4$ for
the standard initial data set with $v_0 = 5$ for $0 \le \theta \le \pi$ for
a simulation containing 20000 spatial grid points in the interval $[0,2 \pi]$.
The peaks in $P$ are essentially the same in that they occur
where $Q,_{\theta} \approx 0$ while the
apparent discontinuity in $Q$ occurs where $\pi_Q \approx 0$ and $P < 0$
($P = 0$ is the horizontal line).}
\end{figure}
The narrow peaks in $P$ occur where
$Q,_{\theta} \approx 0$.
Generically, if $P \approx v \tau$ and $v > 1$, the potential $V_2$
dominates.
The relevant
first integral of (\ref{gowdywave}) is
\begin{equation}
\label{Zintegral}
\left( \frac {dZ}{d \tau} \right)^2 + e^{2Z} Q,_{\theta}^2 = const
\end{equation}
where $Z = P - \tau$. A bounce off $V_2$ yields ${dZ}/{d \tau}  \to -{dZ}/{d
\tau}$ or
$v - 1 \to 1-v$. If the new $v <0$, then $V_1$ dominates yielding the first
integral
\begin{equation}
\label{Pintegral}
P,_{\tau}^2 +\, e^{-2P} \pi_Q^2 = const
\end{equation}
causing $P,_{\tau} \to - P,_{\tau}$ or $v \to -v$.  Eventually, bouncing
between
potentials gives $v < 1$. However, if $Q,_{\theta} \approx 0$, but is not
precisely zero,
it takes a long time for the bounce off $V_2$ to occur.  Precisely at
$\theta_0$
(where $Q,_{\theta} = 0$), $v > 1$ persists.
The apparent discontinuity in $Q$
in Fig.~\ref{gdetail} is not a numerical artifact. It occurs where $P < 0$
and $\pi_Q \approx 0$. Since $Q,_{\tau} = e^{-2P} \pi_Q$ and $\pi_Q \approx c
(\theta - \theta_1)$ (if $\pi_Q = 0$ at $\theta_1$), $Q,_{\tau}$ grows
exponentially in
opposite directions about $\theta_1$. The potential $V_1$ drives $P$ to
positive values unless $\pi_Q = 0$. Thus this feature will narrow as the
simulation proceeds.

Finally, we note that the spatial structure in $P$
scales with the parameter $v_0$ in the initial data. In
our standard initial data set, greater values of $v_0$ lead to the
appearance of
additional spatial structure in a shorter time. The rate of structure formation
decreases and then stops as the AVTD regime is approached.
The scaling is best for the time $\tau_5$
at which the 5th peak appears  in $P$ although it is also seen for $\tau_3$ and
$\tau_7$
(the even nature of the solution causes two peaks to appear at once except at
$\theta = \pi$)
and is described by $\tau_5 = a (v_0 - v_0^{\infty})^{-1}$
where, if $v_0 = v_0^{\infty}$, the 5th peak does not appear until
$\tau = \infty$ for some constant slope $a$.
Explanation of this scaling is still in progress.

Given the success of the symplectic method in studying the singularity
behavior of the Gowdy model, we can consider its extension to the case of
$U(1)$ symmetric cosmologies.
Moncrief has shown \cite{Moncrief86} that cosmological models on $T^3 \times R$
with a spatial $U(1)$ symmetry\index{U(1) symmetric cosmologies}
can be described by five degrees of freedom
$\{ x,z, \Lambda, \varphi, \omega \}$ and their respective conjugate momenta
$\{ p_x, p_z, p_{\Lambda}, p, r \}$.  All variables are functions of spatial
variables $u$, $v$ and time, $\tau$.
If we define a conformal metric $g_{ab}$ in the $u$-$v$ plane as $g_{ab} =
e^{\Lambda} e_{ab}(x,z)$ where
\begin{equation}
\label{eab}
e_{ab} = \frac{1 }{ 2}  \left( \begin{array}{cc}
          e^{2z}+e^{-2z}(1+x)^2 & e^{2z}+e^{-2z}(x^2 - 1) \\
                                                           \\
           e^{2z}+e^{-2z}(x^2 - 1) &  e^{2z}+e^{-2z}(1-x)^2
         \end{array} \right)
\end{equation}
has unit determinant and choose the  conformal lapse $N = e^{\Lambda}$,
Einstein's
equations can be obtained by variation of
\begin{eqnarray}
\label{u1h}
H&=&-\oint {\oint {du\,dv\left\{ {{\textstyle{1
\over 8}}p_z^2+{\textstyle{1 \over
2}}e^{4z}p_x^2+{\textstyle{1 \over 8}}p_{}^2+{\textstyle{1
\over 2}}e^{4\varphi }r_{}^2-{\textstyle{1 \over 2}}p_\Lambda
^2-2p_\Lambda } \right.}} \nonumber \\
& &+\, e^{-2\tau }\left[ {\left( {e^\Lambda e^{ab}} \right),_{ab}-
\;\left( {e^\Lambda e^{ab}} \right),_a\Lambda ,_b+\;e^\Lambda
\left( {e^{-2z}} \right),_{[a}x,_{b]}} \right. \nonumber \\
 & &\left. {\left. {+\, 2e^\Lambda e^{ab}\varphi ,_a\varphi
,_b\, +\, {\textstyle{1 \over 2}}e^\Lambda e^{-4\varphi
}e^{ab}\omega ,_a\omega ,_b} \right]} \right\} \nonumber \\
  &=& H_1 + H_2 .
\end{eqnarray}
Note particularly that
\begin{equation}
\label{u1parts}
H_1 = H_1^G(-2 z,x)+ H_1^G(-2 \varphi, \omega)+H_1^F(\Lambda)
\end{equation}
where $H_1^G(P,Q)$ is the kinetic part of the Gowdy Hamiltonian
(\ref{gowdywaveh}).
Of course, $H_1^F$ is just a free particle Hamiltonian for the degree of
freedom associated with $\Lambda$. This means that not only are the
equations from $H_1$ exactly solvable but also that the Gowdy coding
can be used with essentially no change. The potential term $H_2$ is very
complicated. However, it still contains no momenta so its equations are
trivially exactly solvable. Thus, at least in principle, the extension
of the Gowdy code to the two spatial dimensions of the $U(1)$ code is
completely straightforward.

There are three complications, however, which cause the $U(1)$ problem
to be more difficult. The first involves the initial value problem
(IVP)---the constraints must be satisfied on the initial spacelike slice.
The constraints are
\begin{equation}
\label{u1h0}
{\cal H}_0 = {\cal H} - 2 p_{\Lambda} = 0
\end{equation}
(where ${\cal H}$ is the density in (\ref{u1h})) and
\begin{equation}
\label{u1ha}
H_a=-2 \tilde \pi^b_{a;b}+p_\Lambda \Lambda ,_a-
p_\Lambda ,_a+p\varphi ,_a+r\omega ,_a = 0
\end{equation}
where $\tilde \pi^b_{a}$ is in the 2-space with metric $e_{ab}$
and is linear in $p_x$ and $p_z$ with each term containing one or the other.
Moncrief has proposed a particular solution to the IVP.
First, identically satisfy ${\cal H}_a = 0$ by choosing
\begin{equation}
\label{ivp}
p_x=p_z= \varphi ,_a=\omega ,_a=0 \quad ; \quad p_{\Lambda}
= c \, e^{\Lambda}
\end{equation}
for $c$ a constant. Then, solve ${\cal H}_0 = 0$ for either $r$ or $p$.
Solution is
possible for $c \ge c_{min}$ such that $r^2$ or $p^2 \ge 0$. This allows $x$,
$z$, $\Lambda$, and $p$ or $r$ to be freely specified.
(Without loss of generality, it is possible to set $x = z= 0$ initially
to yield $e_{ab}$ flat. Such a condition may always be imposed at one time
by rescaling $u$ and $v$.)

The second difficulty also involves the constraints. While the Bianchi
identities guarantee the preservation of the constraints by the Einstein
evolution equations, there is no such guarantee for differenced evolution
equations. At this stage of the project, we monitor the maximum value of
the constraints vs $\tau$ over the spatial grid but do nothing else to try
to stay on the constraint hypersurface.

The third difficulty, and the one that is proving to be the greatest obstacle,
is instability associated with spatial differencing in two dimensions. In an
attempt to control the instability, we have introduced a form of third order
accurate spatial
averaging.
In both test cases and generic models, the averaging procedure has
allowed the code to run longer.  However, the averaging can lead to
deviations of the numerical solution from the
correct one. Fortunately, by comparing runs with and without averaging,
these artifacts are easy to identify so that
averaging can allow the code to run long enough for a conclusion about the
asymptotic singularity behavior to be drawn.

Moncrief has provided a test case for the $U(1)$ code. It again starts with
a polarized Gowdy solution and transforms it as either a one-dimensional
($\theta \to u$ or $v$)
or two-dimensional ($\theta \to f(u,v)$) test problem to satisfy the $U(1)$
equations (including the constraints).
As a one-dimensional example, the agreement is excellent and the code can be
run to
large $\tau$. Difficulties arise in the two-dimensional test problem in regions
where the spatial derivatives are large.
In application of the $U(1)$ code to generic models, those which are AVTD
can display nonlinear wave interactions before settling down to $U \to
0$ (where $U$ is the density corresponding to $H_2$ in (\ref{u1h}) and thus
contains all the terms with spatial derivatives),
$z,\varphi, \Lambda \to {\rm const} \  \tau$, and $x,\omega \to$ const.
Increasing spatial resolution will yield narrower (and steeper)
structures and thus may not help to cure instabilities due to steep
gradients.

Despite the limitations of the code,
conclusions can be drawn for generic models in our restricted
class of initial data. We shall consider the models as representative of
subclasses of the data. Models with $r = \omega = 0$ are called polarized.
This condition
is compatible with the above solution to the IVP and is preserved identically
by the
(analytic and numerical) evolution equations. Grubi\u{s}i\'{c} and
Moncrief have
conjectured that these polarized models are AVTD.\cite{bm2}
Therefore, the first model is chosen to be polarized. It exhibits the
conjectured AVTD behavior as shown in Fig.~\ref{polU} for $U$. Detailed
examination of the variables also shows AVTD behavior. Other polarized models
also appear to be AVTD. Actually, this is not too surprising since a polarized
model in our class of initial data must begin with $p_x = p_z = r = \omega =
\varphi,_a = 0$ so that only $\Lambda$, $x$, and $z$ may be freely specified.
But it is possible to set $x = z = 0$ as well without loss of generality
leaving
only $\Lambda$ arbitrary. Thus, polarized models in this class have a very
simple
structure.
\begin{figure}[bth]
\setlength{\unitlength}{1cm}
\makebox[11.7cm]{\psfig{file=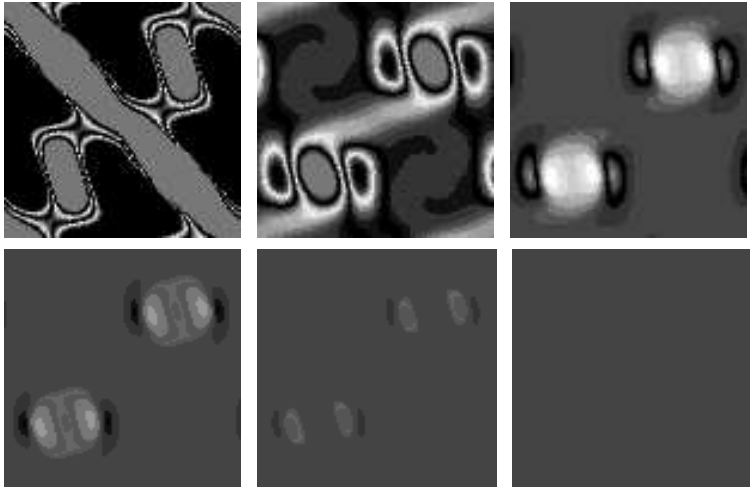,width=10cm}}
\caption
{\protect \label{polU}
Frames of $U(u,v,\tau)$ for the
polarized model $x = z = \Lambda = \sin u \sin v, \  p_\Lambda = 12
e^{\Lambda}, \  \omega = r = 0$.  Time increases to the right and downward.
The final frame corresponds to $U \approx 0$ everywhere.}
\end{figure}
The second model is unpolarized with
$p$ given and the Hamiltonian constraint solved for $r$ in the IVP.
Spatial averaging increases stability and allows the model to be followed
to the point
where only artifacts have $U \ne 0$ as is shown in Fig.~\ref{genU}.
It certainly seems as if this model is also AVTD. Similar behavior is seen in
other examples of this class.
\begin{figure}[bth]
\setlength{\unitlength}{1cm}
\makebox[11.7cm]{\psfig{file=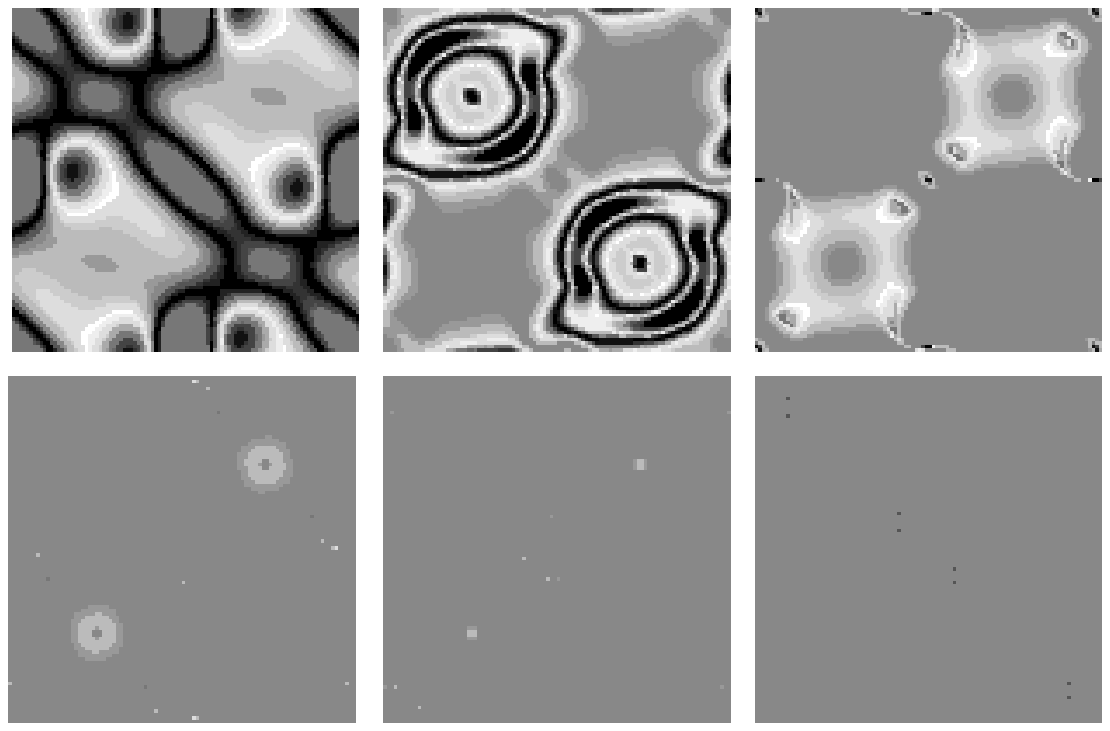,width=10cm}}
\caption[Frames of $U(u,v,\tau)$ for a Generic Model]
{ \protect \label{genU}
Frames of $U(u,v,\tau)$ for the generic model
$x = z = \cos u \cos v , \ \Lambda = \sin u \sin v, \ p_\Lambda =
14 e^{ \Lambda}, \  p = 10 \cos u \cos v$ with averaging.}
\end{figure}
The last
model has $r$ given with $p$ obtained by solving the Hamiltonian constraint
in the IVP. Models of this type are less stable, probably due to the
growth of a steep feature in $\omega$ which does not appear in the other
cases. For this reason, the parameters must be kept small although the
artifacts
still appear more prominently.
Fig.~\ref{genU2} shows
that $U \to 0$ except near artifacts where no statement can be made.
\begin{figure}[bth]
\setlength{\unitlength}{1cm}
\makebox[11.7cm]{\psfig{file=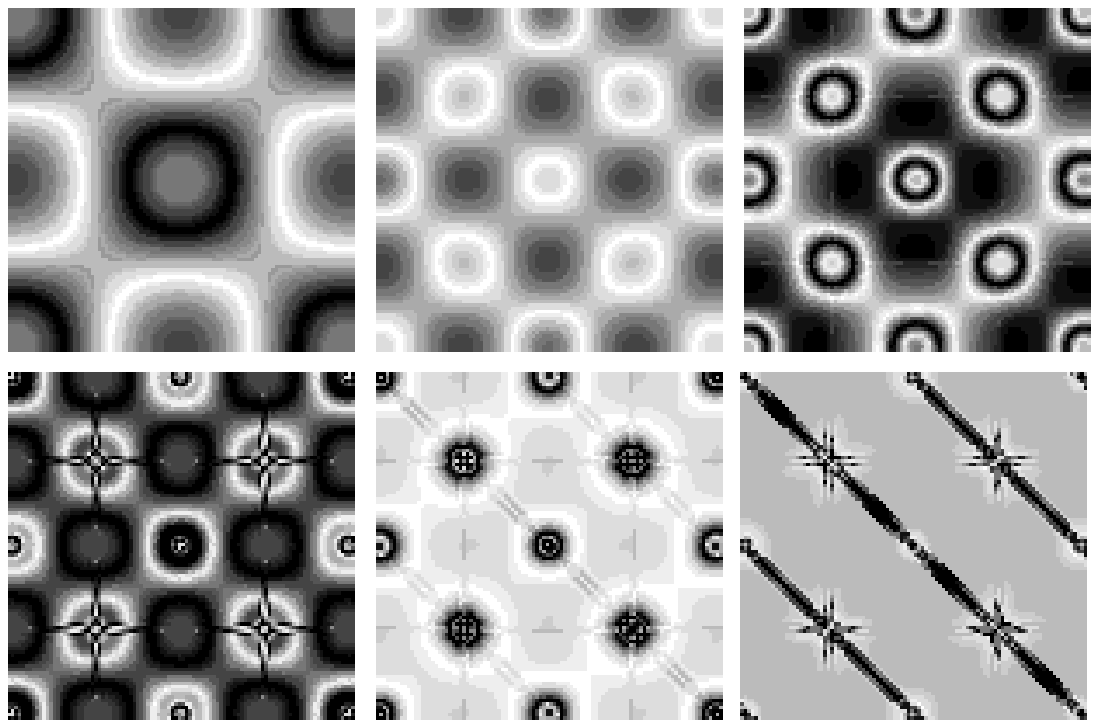,width=10cm}}
\caption[Frames of $U(u,v,\tau)$ for a Second Generic Model]
{\protect \label{genU2}
Frames of $U(u,v,\tau)$ for generic model $x = z = 0,
\  \Lambda = .1 \cos u \cos v, \  p_\Lambda= 2.1 e^
{\Lambda}, \  r = \cos u \cos v$. The diagonal features in the final frames
are numerical artifacts.}
\end{figure}

Thus, we may conclude that the application of the symplectic
method to Einstein's equations for collapsing universes
allows the nature of the singularity to be studied.
Application to the Gowdy model has yielded strong support for its
conjectured AVTD singularity and has allowed the discovery and study
of interesting small scale spatial structure and scaling.
Further progress in understanding the generic singularity of $U(1)$
cosmologies\index{BKL conjecture}
requires imporvements in handling steep spatial gradients.
(Spectral methods to evaluate spatial derivatives have been shown
to work in the Gowdy simulations \cite{bkb95} and may yield improved
behavior in the $U(1)$ case.)
Nevertheless there is strong support that (at least within our restricted
class of initial data) polarized models are AVTD. There is also support
for AVTD behavior in all generic models studied so far. Mixmaster-like
bounces have not been seen while $U$ appears to become small everywhere the
results can be trusted. Several factors could account for
this: (1) the BKL conjecture
might be false; (2) the simulations have not run long enough; (3) Mixmaster
behavior is present but hidden in our variables; or (4) our class of
initial data is insufficiently generic. All these possibilities will be
explored in studies in progress.

\section*{Acknowledgements}
I would like to thank the Astronomy Department of the University of Michigan,
the Institute for Geophysics and Planetary Physics at Lawrence Livermore
National Laboratory, and the Albert Einstein Institute at Potsdam for
hospitality. This work was supported in part by National Science Foundation
Grants PHY93-0559 and PHY9507313. Computations were performed at the
National
Center for Supercomputing Applications (University of Illinois) and at the
Pittsburgh Supercomputer Center.


\section*{References}
\begin{thebibliography}{99}

\bibitem{wald84}
R.M.~Wald, {\it General Relativity}, (University of Chicago, Chicago, 1984)

\bibitem{hawking67}
S.W.~Hawking, {\it Proc.~Roy.~Soc.~Lond.~A}  {\bf 300}, 187 (1967).

\bibitem{he73}
S.W. Hawking, G.F.R. Ellis, {\it The Large Scale Structure of Space-Time}
(Cambridge, Cambridge University, 1973).

\bibitem{hp70}
S.W.~Hawking, R.~Penrose, {\it Proc.~Roy.~Soc.~Lond.~A } {\bf  314}, 529
(1970).

\bibitem{els}
D.~Eardley, E.~Liang, R.~Sachs, {\it J.~Math.~Phys.}~{\bf 13}, 99 (1972).

\bibitem{IM}
J.~Isenberg, V.~Moncrief {\it Ann.~Phys.~(N.Y.)} {\bf 199}, 84
(1990).

\bibitem{bkl71}
V.A.~Belinskii, E.M.~Lifshitz, I.M.~Khalatnikov,
{\it Sov.~Phys.~Usp.}~{\bf 13}, 745--765 (1971).

\bibitem{misner69}
C.W.~Misner, {\it Phys.~Rev.~Lett.}~{\bf 22}, 1071 (1969).

\bibitem{moncriefgr14}
V.~Moncrief, this volume.

\bibitem{rendallgr14}
A.~Rendall, this volume.

\bibitem{finngr14}
L.S.~Finn, this volume.

\bibitem{thorne74}
K.S.~Thorne, in {\it Magic without Magic}, edited by J.~Klauder
(Freeman, San Francisco, 1974).

\bibitem{choptuik93}
M.~Choptuik, {\it Phys.~Rev.~Lett.}~{\bf 70}, 9 (1993)

\bibitem{ch82}
S.~Chandresekhar, J.B.~Hartle,
{\it Proc.~Roy.~Soc.~Lond.}~{\bf A384} 301 (1982).

\bibitem{moser73}
A.R.~Moser, R.A.~Matzner, M.P.~Ryan, Jr., {\it Ann.~Phys.~(N.Y.)}
{\bf 79}, 558 (1973).

\bibitem{rugh90}
S.E.~Rugh, B.J.T.~Jones,  {\it Phys.~Lett.}~{\bf A147}, 353 (1990).

\bibitem{berger94}
B.K.~Berger, {\it Phys.~Rev.~D} {\bf 49}, 1120 (1994).

\bibitem{francisco88}
G.~Francisco, G.E.A.~Matsas, {J.~Gen.~Rel.~ Grav.}~{\bf 20}, 1047
(1988).

\bibitem{burd90}
A.B.~Burd, N.~Buric, G.F.R.~Ellis, {J.~Gen.~Rel.~Grav.}~{\bf 22},
349 (1990).

\bibitem{hobill91}
D.~Hobill, D.~Bernstein, M.~Welge, D.~Simkins, {\it Class.~Quantum
Grav.}~{\bf 8}, 1155  (1991).

\bibitem{pullin91}
J.~Pullin,  in {\it SILARG VII Relativity and Gravitation:  Classical
and Quantum}, 1991.

\bibitem{bt79}
J.D.~Barrow, F.~Tipler, {\it Phys.~Rep.}~{\bf 56}, 372 (1979).

\bibitem{penrose69}
R.~Penrose, {\it Riv.~Nuov.~Cim.}~{\bf 1}, 252 (1969).

\bibitem{shapiro91}
S.L.~Shapiro, S.A.~Teukolsky, {\it Phys.~Rev.~Lett.}~{\bf 66}, 994 (1991).

\bibitem{nakamura82}
T.~Nakamura, H.~Sato, {\it Prog.~Theor.~Phys.}~{\bf 67}, 1396 (1982).

\bibitem{seidel95}
E. Seidel, private communication

\bibitem{wald91}
R.M.~Wald, V.~Iyer, {\it Phys.~Rev.~D} {\bf 44}, 3719 (1991).

\bibitem{tod92}
K.P.~Tod, {\it Class.~Quantum Grav.}~{\bf 9}, 1581 (1992).

\bibitem{shapiro92}
S.L.~Shapiro, S.A.~Teukolsky, {\it Phys.~Rev.~D} {\bf 45}, 2006 (1992).

\bibitem{shapiro93}
S.L.~Shapiro, S.A.~Teukolsky, {\it Astrophys.~J.}~{\bf 419}, 622 (1993).

\bibitem{echeverria93}
F.~Echeverria, {\it Phys.~Rev.~D} {\bf 47}, 2271 (1993).

\bibitem{chiba94}
T.~Chiba, T.~Nakamura, K.~Nakao, M.~Sasaki, {\it Class.~Quantum Grav.}~
{\bf 11}, 431 (1994).

\bibitem{nakamura88}
T.~Nakamura, S.L.~Shapiro, S.A.~Teukolsky, {\it Phys.~Rev.~D} {\bf 38}, 3972
(1988).

\bibitem{wojtkewicz90}
J.~Wojtkiewicz, {\it Phys.~Rev.~D} {\bf 41}, 1867 (1990).

\bibitem{barrabes91}
C.~Barrab\`{e}s, W.~Israel, P.S.~Latlier, {\it Phys.~Lett.}~{\bf 160A},
41 (1991).

\bibitem{barrabes92}
C.~Barrab\`{e}s, A.~Gremain, E.~Lesigne, P.S.~Latlier, {\it Class.
Quantum Grav.}~{\bf 9}, L105 (1992).


\bibitem{hirschmann95a}
E.W.~Hirschmann, D.M.~Eardley, {\it Phys.~Rev.~D} {\bf 51}, 4198 (1995).

\bibitem{abrahams93}
A.M.~Abrahams, C.R.~Evans, {\it Phys.~Rev.~Lett.}~{\bf 70}, 2980 (1993).

\bibitem{garfinkle95}
D.~Garfinkle, {\it Phys.~Rev.~D} {\bf 51}, 5558 (1995).

\bibitem{goldwirth87}
D.S.~Goldwirth, T.~Piran, {\it Phys.~Rev.~D} {\bf 36}, 3575 (1987).

\bibitem{christodoulou87}
D.~Christodoulou, {\it Comm.~Math.~Phys.}~{\bf 105}, 337 (1986);
{\bf 109}, 613 (1987).

\bibitem{hamade95}
R.S.~Hamad\'{e}, J.M.~Stewart, gr-qc/9506044.

\bibitem{evans94}
C.R.~Evans and J.S.~Coleman, {\it Phys.~Rev.~Lett.}~{\bf 72}, 1782 (1994).

\bibitem{hirschmann95b}
E.W.~Hirschmann, D.M.~Eardley, gr-qc/9506078.

\bibitem{koiki95}
T.~Koiki, T.~Hara, S.~Adachi, {\it Phys.~Rev.~Lett.}~{\bf 74}, 5170 (1995).

\bibitem{maison95}
D.~Maison, gr-qc/9504008.

\bibitem{gundlach95}
C.~Gundlach, {\it Phys.~Rev.~Lett.}~{\bf 75}, 3214 (1995).

\bibitem{poisson89}
E.~Poisson, W.~Israel, {\it Phys.~Rev.~Lett.}~{\bf 63}, 1663 (1989).

\bibitem{poisson90}
E.~Poisson, W.~Israel, {\it Phys.~Rev.~D} {\bf 41}, 1796 (1990).

\bibitem{ori91}
A.~Ori, {\it Phys.~Rev.~Lett.}~{\bf 67}, 789 (1991).

\bibitem{ori92}
A.~Ori, {\it Phys.~Rev.~Lett.}~{\bf 68}, 2117 (1992).

\bibitem{gnedin93}
M.L.~Gnedin, N.Y.~Gnedin, {\it Class.~Quantum Grav.}~{\bf 10},
1083 (1993).

\bibitem{brady95}
P.R.~Brady, J.D.~Smith, {\it Phys.~Rev.~Lett.}~{\bf 75}, 1256 (1995).

\bibitem{bonanno}
A.~Bonanno, S.~Droz, W.~Israel, S.M.~Morsink, {\it Proc.~Roy.~
Soc.~Lon.~A} {\bf 450}, 553 (1995).

\bibitem{huebner94}
P. H\"{u}bner, gr-qc/940902.

\bibitem{Fried}
H.~Friedrich, {Comm.~Math.~Phys.}~{\bf 119}, 51 (1988).

\bibitem{cb83}
D.F.~Chernoff, J.D.~Barrow, {\it Phys.~Rev.~Lett.}~{\bf  50}, 134 (1983).

\bibitem{barrow82}
J.D.~Barrow,  {\it Phys.~Rep.}~{\bf 85}, 1 (1982).

\bibitem{khalatnikov85}
I.M.~Khalatnikov, E.M.~Lifshitz, K.M.~Khanin,
L.N.~Shchur, Ya.G.~Sinai, {\it J.~Stat.~Phys.}~{\bf 38}, 97 (1985).

\bibitem{berger91}
B.K.~Berger, {\it J.~Gen.~Rel.~Grav.}~{\bf  23}, 1385 (1991).

\bibitem{ferraz91}
K.~Ferraz, G.~Francisco, G.E.A.~Matsas, {\it Phys.~Lett.}~
{\bf 156A}, 407 (1991).

\bibitem{hobillbook}
D.~Hobill, A.~Burd, A.~Coley, editors, {\it Deterministic Chaos in
General Relativity} (Plenum, New York, 1994).

\bibitem{LKW}
V.G.~LeBlanc, D.~Kerr, J.~Wainwright, {\it Class.~Quantum
Grav.}~{\bf 12} 513 (1995).

\bibitem{BKB95}
B.K.~Berger, ``Comment on the `Chaotic' Singularity in Some Magnetic
Bianchi VI$_0$ Cosmologies,'' submitted to {Class.~Quantum Grav.}

\bibitem{carretero94}
R. Carretero-Gonzalez, H.N. Nunuz-Yepez, A.L. Salas-Brito, {\it Phys.
Lett.} {\bf A188}, 48 (1994).

\bibitem{cotsakis93}
S.~Cotsakis, J.~Demaret, Y.~DeRop, L.~Querella, {\it Phys.~Rev.~D}
{\bf 48}, 4595 (1993).

\bibitem{bkbvm}
B.K.~Berger, V.~Moncrief,
{\it Phys.~Rev.~D} {\bf 48}, 4676 (1993).

\bibitem{bggm95}
B.K.~Berger, D.~Garfinkle, B.~Grubi\u{s}i\'{c}, V.~Moncrief, ``Phenomenology of
the
Gowdy Model on $T^3 \times R$,'' unpublished.

\bibitem{montani}
G.~Montani, {\it Class.~Quantum Grav.}~{\bf 12}, 2505 (1995).

\bibitem{belinskii}
V.A.~ Belinskii, {\it JETP Lett.~}{\bf 56}, 421 (1992).

\bibitem{KK}
A.A.~Kirillov, A.A.~Kochnev, {\it JETP Lett.~}
{\bf 46}, 435 (1987);  A.A.~
Kirillov, {\it JETP} {\bf 76}, 355 (1993).

\bibitem{Moncrief86}
V.~Moncrief,
{\it Ann.~Phys.~(N.Y.)} {\bf 167}, 118 (1986).

\bibitem{gowdy}
R.H.~Gowdy, {\it Phys.~Rev.~Lett.}~{\bf 27} 826 (1971).

\bibitem{bkb74}
B.K.~Berger, {\it Ann.~Phys.~(N.Y.)} {\bf 83}, 458 (1974).

\bibitem{bm1}
B.~Grubi\u{s}i\'{c}, V.~Moncrief,
{\it Phys.~Rev.~D} {\bf 47} 2371 (1993).

\bibitem{fleck}
J.A.~Fleck, J.R.~Morris, M.D.~Feit
{\it Appl.~Phys.}~{\bf 10}, 129 (1976).

\bibitem{vm83}
V.~Moncrief,
{\it Phys.~Rev.~D} {\bf 28}, 2485 (1983).

\bibitem{suzuki}
M.~Suzuki,
{\it Phys.~Lett.}~{\bf A146}, 319 (1990).

\bibitem{bm2}
B.~Grubi\u{s}i\'{c}, V.~Moncrief,
{\it Phys.~Rev. D}~{\bf 49}, 2792 (1994).

\bibitem{bkb95}
B.K.~Berger, ``Application of a Spectral Symplectic Method
to the Numerical Investigation of Cosmological Singularities,''
unpublished.

\end{thebibliography}
\end{document}